\documentclass[12pt]{article}
\usepackage{amsmath,amssymb,mathrsfs,url,graphicx, color}
\usepackage{cite}
\usepackage{booktabs}
\usepackage[hidelinks,pdftex,bookmarks,plainpages=false,pdfpagelabels]{hyperref}
\usepackage{tikz}
\usetikzlibrary{shapes,arrows}

\title{Comparison of the mean field and Bohmian semi-classical approximations to the Rabi model}

\author{Dirk-Andr\'e Deckert,\footnote{Mathematisches Institut, Ludwig-Maximilians-Universit\"at M\"unchen, Theresienstr.\ 39, 80333 M\"unchen, Germany. E-mail: deckert@math.lmu.de} Leopold Kellers\footnote{Mathematisches Institut, Ludwig-Maximilians-Universit\"at M\"unchen, Theresienstr.\ 39, 80333 M\"unchen, Germany. E-mail: bohmian.semiclassical.rabi.model@l.kelle.rs}, Travis Norsen{\footnote{Smith College, USA. E-mail: tnorsen@smith.edu}} and  Ward Struyve\footnote{Instituut voor Theoretische Fysica, KU Leuven, Celestijnenlaan 200D, B--3001 Leuven, Belgium. Centrum voor Logica en Filosofie van de Wetenschappen, K.U.Leuven, Kardinaal Mercierplein 2, B--3000 Leuven, Belgium. E-mail: ward.struyve@gmail.com}
}
\date{}

\def\la{\langle}
\def\ra{\rangle}
\def\pa{\partial}

\def\al{\alpha}

\def\ii{\textrm i}

\newcommand{\be}{\begin{equation}}
\newcommand{\en}{\end{equation}}
\newcommand{\bi}{\begin{itemize}}
\newcommand{\ei}{\end{itemize}}

\begin{document}

\maketitle

\begin{abstract}
\noindent
Bohmian mechanics is an alternative to standard quantum mechanics that does not suffer from the measurement problem. While it agrees with standard quantum mechanics concerning its experimental predictions, it offers novel types of approximations not suggested by the latter. Of particular interest are semi-classical approximations, where part of the system is treated classically. Bohmian semi-classical approximations have been explored before for systems without electromagnetic interactions. Here, the Rabi model is considered as a simple model involving light-matter interaction. This model describes a single mode electromagnetic field interacting with a two-level atom. As is well-known, the quantum treatment and the semi-classical treatment (where the field is treated classically rather than quantum mechanically) give qualitatively different results. We analyse the Rabi model using a different semi-classical approximation based on Bohmian mechanics. In this approximation, the back-reaction from the two-level atom onto the classical field is mediated by the Bohmian configuration of the two-level atom. We find that the Bohmian semi-classical approximation gives results comparable to the usual mean field one for the transition between ground and first excited state. Both semi-classical approximations tend to reproduce the collapse of the population inversion, but fail to reproduce the revival, which is characteristic of the full quantum description. Also an example of a higher excited state is presented where the Bohmian approximation does not perform so well.
\end{abstract}

\section{Introduction}
Bohmian mechanics is an alternative to standard quantum theory that describes the motion of actual point-particles whose motion is guided by the wave function.
While Bohmian mechanics reproduces the usual quantum predictions (insofar the latter are unambiguous), it offers novel ways to solve practical problems \cite{burghardt04,burghardt05,wyatt05,oriols07,albareda09,poirier10,parlant12,benseny14,oriols19}.
For example, interacting quantum systems are often hard to analyse analytically or even numerically.
Therefore one often resorts to approximation methods.
One such approach is the semi-classical approximation (sometimes also called mixed quantum-classical approximation), where part of the system is treated classically.
The simplest way is by means of the mean field approximation.
In this case, the back-reaction is through an average force, where the average is taken over the quantum system, which acts on the classical system.
Consider for example the interaction between a quantum particle and a classical particle.
The quantum particle is described by a wave function $\chi({{\bf x}}_1,t)$ which satisfies the Schr\"odinger equation{\footnote{We will always denote the actual position and configuration by capital letters and the arguments of the wave function by small letters.}}
\begin{equation}
\ii \hbar \pa_t \chi({{\bf x}}_1,t) =  \left[ - \frac{\hbar^2}{2m_1}\nabla^2_1  + V\left({\bf x}_1,{{\bf X}}_2(t)\right) \right] \chi({{\bf x}}_1,t),
\label{0.1}
\end{equation}
where the potential $V$ is evaluated at the position of the second particle ${\bf X}_2$.
The second particle satisfies Newton's equation
\begin{align}
m_2 {\ddot {{\bf X}}}_2(t) &=   -\left\langle \chi \left| {\boldsymbol \nabla}_2   V\left({\bf x}_1,{{\bf X}}_2(t)\right) \right| \chi \right\rangle \nonumber\\
&= \int d^3x_1\left|\chi({{\bf x}}_1,t)\right|^2  \left[-{\boldsymbol \nabla}_2 V\left({{\bf x}}_1,{{\bf X}}_2(t)\right)\right],
\label{0.2}\end{align}
where the force is averaged over the state of the quantum particle.

In the Bohmian semi-classical approximation \cite{gindensperger00,prezhdo01,struyve20a}, \eqref{0.2} is replaced by
\begin{equation}
m_2 {\ddot {{\bf X}}}_2(t) =   -   {\boldsymbol \nabla}_2   V({{\bf X}}_1(t),{{\bf X}}_2(t)) ,
\label{0.4}
\end{equation}
where ${{\bf X}}_1$ is the actual position of particle one, which satisfies the Bohmian guidance equation 
\begin{equation}
{\dot {{\bf X}}}_1(t) = {\boldsymbol v}^\chi({{\bf X}}_1(t),t) ,
\label{0.3}
\nonumber\end{equation}
where
\begin{equation}
{\boldsymbol v}^\chi = \frac{\hbar}{m_1} {\textrm{Im}} \frac{{\boldsymbol \nabla} \chi}{\chi} ,
\nonumber\end{equation}
and where $\chi$ satisfies the Schr\"odinger equation \eqref{0.1}.
So the classical particle is not acted upon by some average force, but rather by the actual particle of the quantum system.
The Bohmian semi-classical approximation has been studied for a number of different systems \cite{gindensperger00,prezhdo01,gindensperger02a,gindensperger02b,meier04,garaschuk11}, leading to results that are often close to those predicted by full quantum theory and often performing better than the mean field approach.
Higher order corrections to the Bohmian semi-classical approximation are considered in \cite{oriols07,norsen15,oriols19}.

The Bohmian semi-classical approximation hence offers a promising novel approach, which could potentially be useful in other domains, such as quantum field theory or quantum gravity (see \cite{struyve20a,struyve17a} for further development in these domains).
As there is no agreed upon proposal for a quantum theory of gravity and it is not even clear whether gravity should be quantized at all, semi-classical methods may provide a good toolbox to find possible effects of quantum gravitational origin even before settling for the fundamental theory of quantum gravity.
In order to manage our expectations, however, it is instructive to consider other situations in the realm of non-relativistic quantum mechanics, where we know the exact quantum results and where the Bohmian semi-classical approximation can be tested.
This will guide us in distilling which aspects of the semi-classical approach can and which cannot be taken as aspects of the full quantum description.
While the Bohmian semi-classical approximation has so far been applied for numerical computations of scattering, the purpose of this paper is to consider a simple model in the completely different context of quantum optics, namely that of the well-known Rabi model.
As such it also forms an exploration into the Bohmian treatment of electromagnetism, which has been little considered before \cite{bohm52b,bohm87,kaloyerou94,kaloyerou03,struyve10}.

We start with outlining the Rabi model in section \ref{rabi}, followed by the mean field and the Bohmian semi-classical approximations in sections \ref{usa} and \ref{bsa} respectively.
Then we compare the two approximations in section \ref{comparison}.
We consider transitions both between the 1s and 2p states for the hydrogen atom, with weak as well as strong coupling, and transitions between the 1s and 9p states.
We conclude in section \ref{conclusion}.

\section{Rabi model}\label{rabi}
The Rabi model describes a single electromagnetic field mode interacting with a two-level atom \cite{scully97}.
We start with reviewing this model and its properties here. In the case of a single electron of mass $m$ in a hydrogen atom and assuming the dipole approximation (appropriate when the field wavelength $1/\nu$ is larger than the atomic size), the full quantum Hamiltonian is the sum of the free atom Hamiltonian, the free field Hamiltonian and the interaction Hamiltonian:
\be
{\widehat H} = {\widehat H}_A + {\widehat H}_F + {\widehat H}_I,
\label{1}
\en
where
\be
{\widehat H}_A = \frac{{\widehat {\bf p}}^2}{2m} + {\widehat V} , \qquad {\widehat H}_F = \hbar \nu \left( {\widehat a}^{\dagger} {\widehat a} + \frac{1}{2}\right), \qquad {\widehat H}_I = -e {\widehat {\bf x}} \cdot {\widehat {\bf E}} ,
\en
with
\be
{\widehat {\bf E}} = {\bf e}_p {\mathcal E} ({\widehat a} + {\widehat a}^{\dagger})
\en
the electric field operator. ${\bf e}_p$ is the unit polarization vector and ${\mathcal E} = \sqrt{\frac{\hbar \nu}{2 \epsilon_0 {\mathcal V}}} $ with ${\mathcal V}$ the volume of the cavity.

Choosing the position representation for the electron and the field, where the latter is defined by
\be
{\widehat a} = \sqrt{\frac{\nu}{2\hbar}} \left( q + \frac{\hbar}{\nu} \frac{\pa}{\pa q} \right), \qquad {\widehat a}^{\dagger} =  \sqrt{ \frac{\nu}{2\hbar}} \left( q - \frac{\hbar}{\nu} \frac{\pa}{\pa q}\right),
\label{eq:position_translation}
\en
the total Hamiltonian reads
\be
{\widehat H} = -\frac{\hbar^2}{2m}\nabla^2 + V({\bf x}) - \frac{\hbar^2}{2} \frac{\pa^2}{\pa q^2} + \frac{\nu^2}{2} q^2 + \alpha q {\bf x} \cdot {\bf e}_p ,
\label{10}
\en
where
\be
\alpha =  - e {\mathcal E} \sqrt{\frac{2\nu}{\hbar}},
\en
and which acts on wave functions $\psi=\psi({\bf x}, q)$. In this representation
\be
{\widehat {\bf E}} = - {\bf e}_p \frac{\alpha}{e} q .
\en

For a two-level atom, we only need to consider superpositions of two energy eigenstates $|+\ra$ and $|-\ra$, given by $\phi_\pm({\bf x}) = \la {\bf x} |\pm \ra$ in the position basis, which have energies $E_+$ and $E_-$.
We assume that the atom is in resonance with the field, i.e., $E_+ - E_- = \hbar \nu$.
As usual, we further assume that 
\be
\langle + | {\bf x} | +\rangle = \langle  -| {\bf x} |- \rangle = 0, \qquad \langle + | {\bf x} | - \rangle = \langle - | {\bf x} | + \rangle=  \int d^3x \phi^*_+({\bf x}){\bf x} \phi_-({\bf x}),
\en
and write ${\mathcal P} ={\bf e}_p\cdot  \langle + | {\bf x} | - \rangle $, which is the transition dipole moment in atomic units.
In the energy basis for the atom, we have 
\be
\psi = C_+(q,t) |+\ra + C_-(q,t) |-\ra
\label{15}
\en
and 
\be
{\widehat H}_A = E_+ |+\ra  \la+| + E_-|-\ra  \la -| ,\qquad {\widehat H}_I =  \hbar g  ({\widehat a} + {\widehat a}^{\dagger})  \big( |+\ra  \la-| + |-\ra  \la +|\big),
\label{eq:two_level_hamiltonian_terms}
\en
where
\be
g = - \frac{e{\mathcal E}   {\mathcal P} }{\hbar} = \al \frac{{\mathcal P} }{\sqrt{2\nu \hbar}}.
\label{eq:coupling}
\en
The Schr\"odinger dynamics can then be written as 
\be
 {\dot C}_+ = - \frac{\ii}{ \hbar}\left( E_+ C_+ + {\widehat H}_F C_+  + \al  q {\mathcal P} C_-\right) , \qquad  {\dot C}_- = - \frac{\ii}{ \hbar}\left( E_- C_-  + {\widehat H}_F C_- + \al  q {\mathcal P} C_+\right). 
\en
Using the photon number basis $|n\ra$ for the electromagnetic field mode, the state \eqref{15} reads
\be
\psi = \sum_n \left( C_{n,+}(t) |+\ra|n\ra + C_{n,-}(t) |-\ra|n\ra \right).
\label{16}
\en
The observable of interest is the population inversion
\be
W(t) = \sum_n \left( |C_{n,+}(t)|^2 - |C_{n,-}(t)|^2 \right).
\en
The population inversion obtains values $\pm 1$ whenever the atom has energy $E_\pm$.

This dynamics can be explicitly solved in the rotating phase approximation, which is valid for $g \ll \nu$, and which amounts to dropping the energy non-conserving terms in the interaction Hamiltonian, i.e., 
\be
{\widehat H}_I \approx  \hbar g  \big( {\widehat a}|+\ra  \la-|  + {\widehat a}^{\dagger}  |-\ra  \la +|\big).
\label{eq:jaynes_cummings_interaction}
\en
This approximate Hamiltonian (together with ${\widehat H}_A$ and ${\widehat H}_F$ in the same representation) is called the Jaynes--Cummings model. If the state is initially in the excited state, i.e.\ $C_{n,-}(0)=0$, then the population inversion is
\be
W(t) = \sum_n  \rho_n(0) \cos(\Omega_n t)  
\en
with $\rho_n(0) =|C_{n,+}(0)|^2$ denoting the probability that there are $n$ photons present at time $t=0$ and 
\be
\Omega_n = 2|g| \sqrt{n+1}
\en
the Rabi frequency for the $n$-particle sector.

In the Bohmian theory, there is an actual position ${\bf X}$ for the electron and a configuration $Q$ for the field mode
{\footnote{The Bohmian treatment of the Rabi model follows from the Bohmian treatment of electrodynamics where a field configuration is introduced for the electromagnetic field  \cite{bohm52b,bohm87,kaloyerou94,struyve10}.}}.
They satisfy the guidance equations
\be
\dot  {\bf X}(t) = \frac{\hbar}{m} {\textrm{Im}}\left(\frac{{\boldsymbol \nabla}\psi({\bf x}, q,t) }{\psi({\bf x}, q,t)}\right) \Bigg|_{{\bf x} = {\bf X}(t), q = Q(t)} , 
\label{eq:guidance_X}
\en
\be
\dot Q(t) = \hbar {\textrm{Im}}\left(\frac{1 }{\psi({\bf x}, q,t)} \frac{\partial \psi({\bf x}, q,t)}{\partial q}\right) \Bigg|_{{\bf x} = {\bf X}(t), q = Q(t)},
\label{eq:guidance_Q}
\en
where $\psi({\bf x} ,q,t)$ satisfies the Schr\"odinger equation with Hamiltonian \eqref{10}.
Differentiation in time yields the Newtonian-like equations
\be
m \ddot {\bf X}(t) = - \alpha Q(t) {\bf e}_p -{\boldsymbol \nabla} (V({\bf x})   + Q^\psi({\bf x} ,q,t))\Big|_{{\bf x} = {\bf X}(t), q = Q(t)}  ,\quad 
\en
\be
 \ddot Q(t) + \nu^2 Q(t) = - \alpha {\bf X}(t) \cdot {\bf e}_p   - \frac{\pa Q^\psi({\bf x} ,q,t)}{\pa q}     \Bigg|_{{\bf x} = {\bf X}(t), q = Q(t)} ,
\label{90}
\en
where
\be
Q^\psi= - \frac{\hbar^2}{2m} \frac{\nabla^2 |\psi|}{|\psi|}- \frac{\hbar^2}{2} \frac{1}{|\psi|}\frac{\pa^2|\psi|}{\pa q^2}
\en
is the quantum potential.

By virtue of equations \eqref{eq:guidance_X} and \eqref{eq:guidance_Q} the distribution $|\psi({\bf X},Q,t)|^2$ is preserved over time.
This property is referred to as {\em equivariance} \cite{duerr09}.
With this distribution, called the {\em quantum equilibrium distribution}, the usual quantum mechanical predictions are obtained.  

In order to compare the semi-classical approximations to the full quantum treatment, we will consider the initial product state 
\be
\psi({\bf x},q,0) = \chi({\bf x}) \psi_\gamma(q)
\label{19}
\en
at time $t=0$, where 
\be
\psi_\gamma = \left(\frac{\nu}{\pi \hbar}  \right)^{1/4} \exp\left[ - \frac{\nu}{2 \hbar} \left(q- \gamma \sqrt{\frac{2\hbar}{\nu}} \right)^2 \right]
\label{20}
\en
is a coherent state for the field mode, parameterized by $\gamma = \gamma_r + \ii \gamma_i$ ($\gamma_r$ and $\gamma_i$ real), with the average number of photons equal to $\la {\widehat N} \ra = |\gamma|^2$.
For this state, the initial photon probability distribution is \be
\rho_n(0) = \frac{\la {\widehat N} \ra^n \exp^{-\la {\widehat N} \ra}}{n!} .
\en
This state is interesting because the time evolution of $W(t)$ shows the usual Rabi oscillations (with frequency of the order $\Omega_{\la {\widehat N} \ra}$ when $\la {\widehat N} \ra \gg 1$) and the usual collapse of the amplitude of the oscillation (after time $t_c \simeq 1/2g$), but unlike other initial states there is also a revival of the amplitude occurring after times $t_r \simeq 2\pi \sqrt{\la {\widehat N} \ra}/g$ \cite[p.\ 201]{scully97}. 

In the Bohmian semi-classical approximation, to be discussed in section \ref{bsa}, we will need an initial distribution for the initial position ${\bf X}$ and the field configuration $Q$ and its velocity $\dot Q$. We will take it to be the one implied by the full quantum theory discussed above, without any semi-classical approximation. The resulting distribution is $\rho_a({\bf X})\rho_f(Q,\dot Q)$, with 
\be
\rho_a({\bf X}) = |\chi({\bf X})|^2 
\label{21}
\en
and
\begin{align}
\rho_f(Q,\dot Q ) &=  |\psi_\gamma(Q)|^2 \delta\left(\dot Q - \sqrt{2 \nu \hbar } \gamma_i \right) \nonumber\\
&=   \sqrt{\frac{\nu}{\pi \hbar}} \exp\left[ - \frac{\nu}{ \hbar} \left(Q- \gamma_r \sqrt{\frac{2\hbar}{\nu}} \right)^2 \right]  \delta\left(\dot Q - \sqrt{2 \nu \hbar } \gamma_i \right) ,
\label{22}
\end{align}
In the mean field semi-classical approximation, we will merely need the initial distribution of $Q$ and $\dot Q$, which we will take to be $\rho_f(Q,\dot Q )$.

\section{Usual mean field semi-classical approximation}\label{usa}
In this semi-classical approximation, the electromagnetic field mode evolves classically, while the atom is described quantum mechanically.
The coupled equations of motion read 
\be
\ii \hbar \frac{\pa \chi({\bf x}, t)}{\pa t} = \left[ -\frac{\hbar^2}{2m}\nabla^2 + V({\bf x}) + \alpha Q(t) {\bf x} \cdot {\bf e}_p \right]\chi({\bf x}, t),
\label{100}
\en
\be
{\ddot Q}(t) + \nu^2 Q(t) = - \alpha \langle {\bf x} \rangle \cdot {\bf e}_p ,
\label{101}
\en
where 
\be
\langle {\bf x} \rangle = \int d^3 x |\chi({\bf x}, t)|^2 {\bf x}.
\label{102}
\en
In the case of the two-level atom, the wave function of the electron is a superposition of two energy eigenstates
\be 
\chi({\bf x},t) = C_+(t) \phi_+({\bf x}) + C_-(t) \phi_-({\bf x}),
\label{103}
\en
so that the equations of motion reduce to
\be
 {\dot C}_+ = - \frac{\ii}{ \hbar}\left( E_+ C_+ + \al  Q {\mathcal P} C_-\right) , \qquad  {\dot C}_- = - \frac{\ii}{ \hbar}\left( E_- C_- + \al  Q {\mathcal P} C_+\right) ,
\label{104}
\en
\be
{\ddot Q} + \nu^2 Q = - 2 \alpha {\mathcal P} \textrm{Re}(C^*_+C_-) .
\label{105}
\en
Note that the wave function should be normalized to one. So in particular we have
\be
 |C_+|^2 + |C_-|^2  = 1.
\label{105.1}
\en
The normalization is preserved by the dynamics.

The population inversion can be defined similarly as before. Given initial data $Q(0),\dot Q(0), \chi({\bf x},0)$, there is a unique solution to the dynamical equations. For such a solution the population inversion is
\be
W(t) = |C_+(t)|^2 - |C_-(t)|^2.
\label{106}
\en
For an ensemble, with different initial field configurations, the population inversion should be averaged.
For example, we will consider the semi-classical approximation corresponding to a coherent state for the field mode.
This means that we will average over the distribution \eqref{22}.

When ignoring the back-reaction of the atom onto the field (which corresponds to taking the right-hand-side of \eqref{105} zero), a solution for the electric field is given by  
\be
{\bf E}(t) = {\mathcal E} \cos(\nu t)  {\bf e}_p  \qquad {\textrm{or}} \qquad       Q(t) = \sqrt{\frac{\hbar}{2\nu}} \cos(\nu t).
\en
Using the rotating phase approximation (for $g \ll \nu$), the population inversion 
\be
W(t)  = |C_+(t)|^2 - |C_-(t)|^2= \cos(\Omega_R t)
\en
is periodic with Rabi frequency $\Omega_R = |g|$.

\section{Bohmian semi-classical approximation}\label{bsa}
In the Bohmian semi-classical approximation, the wave equation for the atom is the same as in the mean field semi-classical approximation. The key difference is that the classical field mode interacts with the actual Bohmian position of the atom. The dynamics is
\be
\ii \hbar \frac{\pa \chi({\bf x}, t)}{\pa t} = \left[ -\frac{\hbar^2}{2m}\nabla^2 + V({\bf x}) + \alpha Q(t) {\bf x} \cdot {\bf e}_p \right]\chi({\bf x}, t),
\label{eq:semiclassical_quantum_part_schroedinger}
\en
\be
{\dot {\bf X}}(t) = \frac{\hbar}{m} \textrm{Im}\left( \frac{{\boldsymbol \nabla} \chi({\bf x}, t)}{\chi({\bf x}, t)}\right) \Big|_{{\bf x} = {\bf X}(t)}, \qquad {\ddot Q}(t) + \nu^2 Q(t) = - \alpha  {\bf X}(t)  \cdot {\bf e}_p .
\label{eq:semiclassical_classical_part_Bohmian}
\en
This approximation can be derived from the full Bohmian theory, see \cite{struyve20a} for details. (The wave equation \eqref{eq:semiclassical_quantum_part_schroedinger} is obtained as an approximation to the equation for the conditional wave function for the atom. The equation for $Q$ is obtained by ignoring the quantum potential in \eqref{90}.)

For a two-level atom with wave function \eqref{103}, these equations reduce to 
\be 
 {\dot C}_+ = - \frac{\ii}{ \hbar}\left( E_+ C_+ + \al  Q {\mathcal P} C_-\right) , \qquad  {\dot C}_- = - \frac{\ii}{ \hbar}\left( E_- C_- + \al  Q {\mathcal P} C_+\right) ,
\label{eq:atom_eom_semiclassical}
\en
\be
 {\dot {\bf X}} = \frac{\hbar}{m} \textrm{Im}\left( \frac{ C_+ {\boldsymbol \nabla} \phi_+ + C_- {\boldsymbol \nabla} \phi_-}{C_+ \phi_+ + C_- \phi_-}\right) , \qquad  {\ddot Q}(t) + \nu^2 Q(t) = - \alpha  {\bf X}(t)  \cdot {\bf e}_p .
\label{eq:eom_field_guiding_Bohmian}
\en
This dynamics also preserves the normalization \eqref{105.1}.

Given initial data ${\bf X}(0), Q(0),\dot Q(0), \chi({\bf x},0)$, there is a unique solution to the dynamical equations.
The population inversion, which can be defined as in \eqref{106}, depends on all the initial data.
Therefore for an ensemble, the population inversion should be averaged.
For example, in the case of the coherent state for the field mode we will average the population inversion over the position and field distributions \eqref{21} and \eqref{22}.

Actually, this definition of the population inversion might not correspond to the measured one. Namely, from a fundamental point of view, in Bohmian mechanics the outcome of any measurement of an observable depends on the actual positions of the particles. In the case of the full theory, equivariance implies that if the atomic energy eigenstates have negligible overlap, then the position of the particle tends to correlate with the state of the atom, because it will be in the support of either one of the energy eigenstates. However, this latter property does not hold any longer in the semi-classical approximation, because of the lack of equivariance. Therefore, the populations inversion should not be inferred from the wave function in this case. However, we will not attempt a more correct approach, since this would require an actual modeling of the measurement, which would take us too far.

The Bohmian semi-classical approximation has been compared to the mean field one in \cite{prezhdo01} for non-relativistic particle scattering.
These approximations were computationally equally demanding.
In the mean field approximation, at each time step an average force needs to be calculated (by averaging over the quantum particle).
On the other hand, in the Bohmian case, different initial position of the quantum particle need to be considered, leading to different wave function evolutions. 
In the case we are considering here, the Hilbert space of the quantum particle is two-dimensional and the induced force on the classical system is simply computed (no numerical integration is necessary).
In the Bohmian case, the trajectories need to be calculated for the quantum particles and therefore this is computationally more demanding. Note that there is a priori no relation between the two approximations. It is for example not so that the mean field approximation follows from the Bohmian one by some averaging.

The following difference can already be noted between the two semi-classical approximations.
In the full quantum theory, there may be spontaneous emission, which is absent in the mean field semi-classical approximation \cite{scully97}.
That is, according to the full quantum description, an excited atom may make the transition to the lower state if the field mode starts in the vacuum.
In the semi-classical approximation, taking the vacuum to be determined by a field that is zero initially, i.e., $Q(0)= \dot Q (0) = 0$, we have that the initial conditions $C_2(0) = 1$, $C_1(0)= 0$ yield the solution $C_2(t) = \exp(-\ii E_2t)$, $C_1(t)=Q(t) = 0$, so that this transition can not be made.
In the Bohmian case, however, such a transition does happen.
Namely, the evolution of $Q$ then also depends on the initial position ${\bf X}(0)$.
The field mode only stays zero in the special case when ${\bf X}(t) = 0$. This happens for example in the special case where the phase of $\phi_2$ does not depend on the spatial coordinates and the initial position is ${\bf X}(0)=0$.
 So for a generic ${\bf X}(0)$ (distributed according to $|\psi|^2$), there will be a transition.
A similar remark applies when the atom starts in the ground state and the field mode is zero.
According to the mean field approximation, the atom does not get excited.
But it typically does get excited in the Bohmian approximation.
This being said, taking $Q(0)= \dot Q (0) = 0$ in the semi-classical approximation does not really characterize the vacuum state.
A better approximation consists of taking a distribution over $Q(0)$ and $ \dot Q (0)$ (as we are doing in the case of the coherent state).
This would then take into account also the ``vacuum fluctuations'', i.e., initial fields which are not zero, and for those there is spontaneous emission.

\section{Comparison of the semi-classical approaches}\label{comparison}

\begin{figure}
        \centering
        \tikzset{box/.style = {draw, shape=rectangle,
        text width=10em, text centered, minimum height=7em},
        line/.style = {draw, very thick, -latex'},
        solution/.style = {draw, shape=ellipse, align=center, text width=7em, node distance=3cm, minimum height=3em}
}
\resizebox{\textwidth}{!}{
        \begin{tikzpicture}[node distance = 6cm, auto]
                \node [box] (pauli) {minimal-coupling Hamiltonian};
                \node [box, right of=pauli] (dipole) {dipole Hamiltonian \eqref{1}};
                \node [box, below of=pauli, fill=black!50!green, node distance=11cm] (twolevelquantum) {2-level dipole Hamiltonian \eqref{eq:two_level_hamiltonian_terms}};
                \node [box, below of=twolevelquantum, fill=black!50!green] (jaynescummings) {Jaynes--Cummings Hamiltonian \eqref{eq:jaynes_cummings_interaction}};
                \node [box, right of=dipole] (position) {dipole Hamiltonian in position representation \eqref{10}};
                \node [box, below of=dipole] (semiclassical1) {Mean field SCA \eqref{100}-\eqref{102}};
                \node [box, right of=semiclassical1] (semiclassical2) {Bohmian SCA \eqref{eq:semiclassical_quantum_part_schroedinger},\eqref{eq:semiclassical_classical_part_Bohmian} };
                \node [box, below of=semiclassical1, fill=blue, node distance=5cm] (twolevelsemiclassical1) {Mean field SCA for two-level atom \eqref{104}, \eqref{105}};
                \node [box, below of=semiclassical2, fill=red, node distance=5cm] (twolevelsemiclassical2) {Bohmian SCA for two-level atom \eqref{eq:atom_eom_semiclassical}, \eqref{eq:eom_field_guiding_Bohmian}};
                \path [line] (pauli) -- node[align=center] {dipole} (dipole);
                \path [line] (pauli) -- node[align=center, below] {approxi-\\mation} (dipole);
                \path [line] (dipole) -- ++(0,-2cm) -| ++(-6cm,-3.5cm)-- node[align=left, below right] {resonance:\\  restriction to \\ 2-level atom} (twolevelquantum);
                \path [line] (twolevelquantum) -- node[align=left] {weak coupling:\\rotating wave\\approximation} (jaynescummings);
                \path [line] (dipole) -- node {\eqref{eq:position_translation}} (position);
                \path [line] (position) -- node {} (dipole);
                \path [line] (position) -- ++(0,-4cm) -| (semiclassical1);
                \path [line] (position) -- node[align=right, left] {semi-classical approximation (SCA)} (semiclassical2);
                \path [line] (semiclassical1) -- node[align=left] {restriction to \\ 2-level atom}  (twolevelsemiclassical1);
                \path [line] (semiclassical2) -- node[align=left] {restriction to \\ 2-level atom}  (twolevelsemiclassical2);
        \end{tikzpicture}
}
\caption{Overview of the different theories involved and the approximations that are being made. The colors match those of the plots. The results for the semi-classical approximations are obtained numerically and do not feature the rotating wave approximation. They are compared to solutions of the full quantum case, which are of numerical origin unless the rotating wave approximation applies, in which case there is a known explicit solution to the Jaynes-Cummings Hamiltonian.}
\label{fig:derivation}
\end{figure}
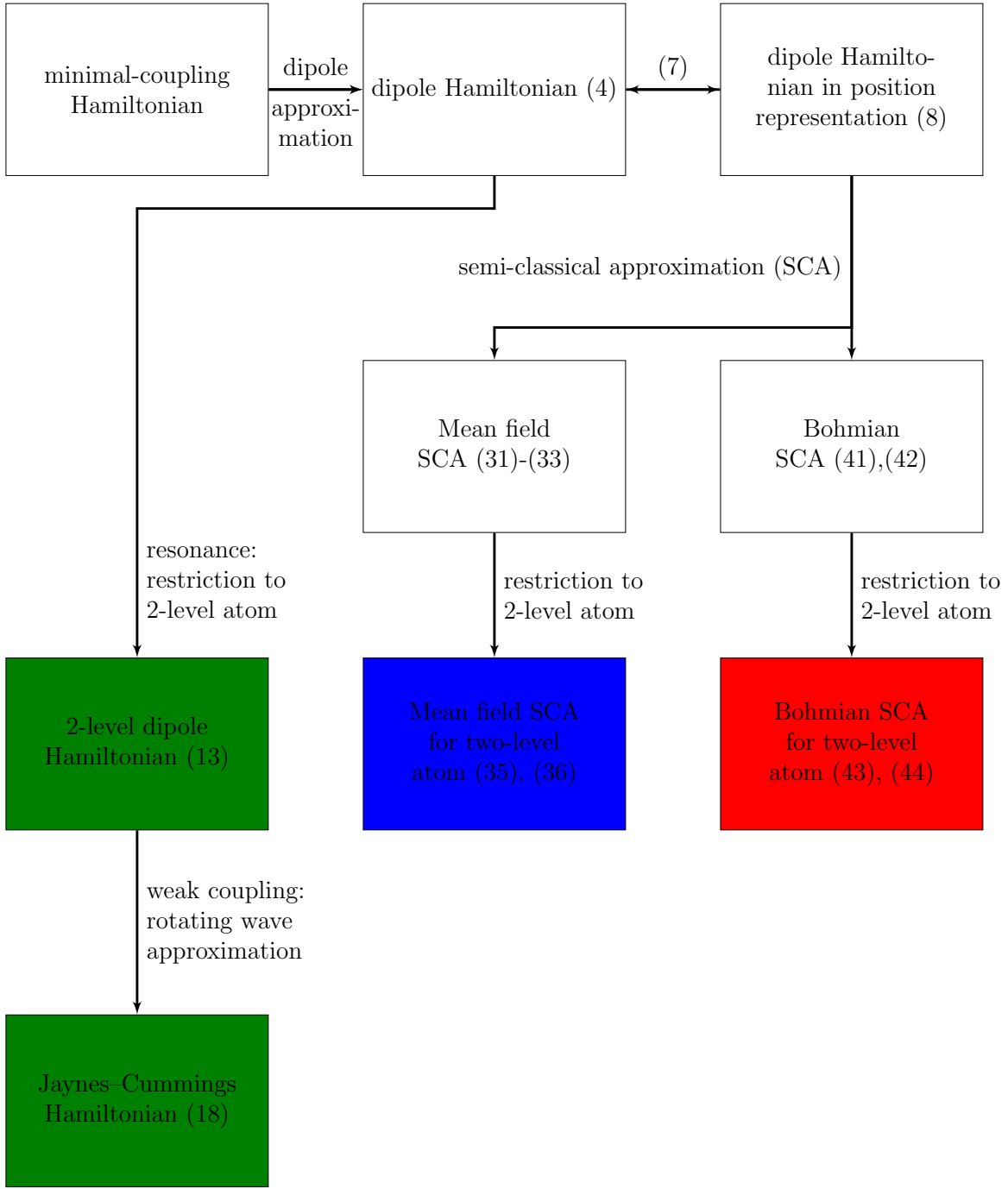
We will now compare the usual mean field and the Bohmian semi-classical approximation to the exact quantum solution of the Jaynes--Cummings model.
Figure \ref{fig:derivation} visually summarizes the previous chapters' derivation of the equations to be solved here.

For the exact quantum result, we assume the initial wave function \eqref{19}, with the initial coherent state \eqref{20} for the field mode.
We further consider the atom to be the hydrogen atom with energy eigenstates $\psi_{n,m,l}$.
Choosing\footnote{So the electron neither has angular momentum nor spin in this treatment. This is sufficient as adding them turns out to merely increase computational effort without having a qualitative impact on the results presented here.} the states $|+ \ra = \psi_{n_+,l_+,m_+}$ and $|- \ra = \psi_{n_-,l_-,m_-}$ and taking ${\bf e}_p$ in the $z$-direction fixes ${\mathcal P}=\langle +|z| -\rangle$ and $\nu=(E_+ - E_-)/\hbar$.
This leaves us free to choose $\al$, which determines $g$ via equation \eqref{eq:coupling}.
Unless indicated otherwise, all numbers will be given in atomic units, where $\hbar = m_\text{e}=e=1$ and $\epsilon_0=4\pi$.

We will consider transitions between the 1s and 2p states, both with weak coupling (where the full quantum result will be derived using the rotating wave approximation) and beyond, as well as transitions between the 1s and 9p states. The initial wave function (at $t=0$) is assumed to be either the excited or the ground state. For the semi-classical approximations, the initial configurations are random, with probability distribution given by $\rho_a({\bf X})\rho_f(Q,\dot Q)$ in the Bohmian case and by $\rho_f(Q,\dot Q)$ in the mean field case, with $\rho_a$ and $\rho_f$ given in \eqref{21} and \eqref{22}. We take $\gamma\in\mathbb{R}$ so that the initial field satisfies $\dot Q=0$, so that the expected number of photons is $\langle\widehat N\rangle = \gamma^2$.

While in the full quantum case the population inversion has a precise numerical value, calculating it in the semi-classical approaches however means averaging the
distinct values over an ensemble. It is thus desirable to be able to have grips on the
quality of the statistics. In order to achieve this, the ensemble with a total of 2500 initial
conditions is arbitrarily split into 5 batches of size 500 each. The variance among the
averages for each of those batches can then be used to derive confidence intervals for the
total mean. With such confidence intervals shown as shaded areas around the individual
graphs presented in the following subsections, we conclude that the respective results
are actually statistically meaningful.

\subsection{Transitions between 1s and 2p state, with weak coupling}
\begin{figure}[p]
        \centering
        \includegraphics[width=\textwidth]{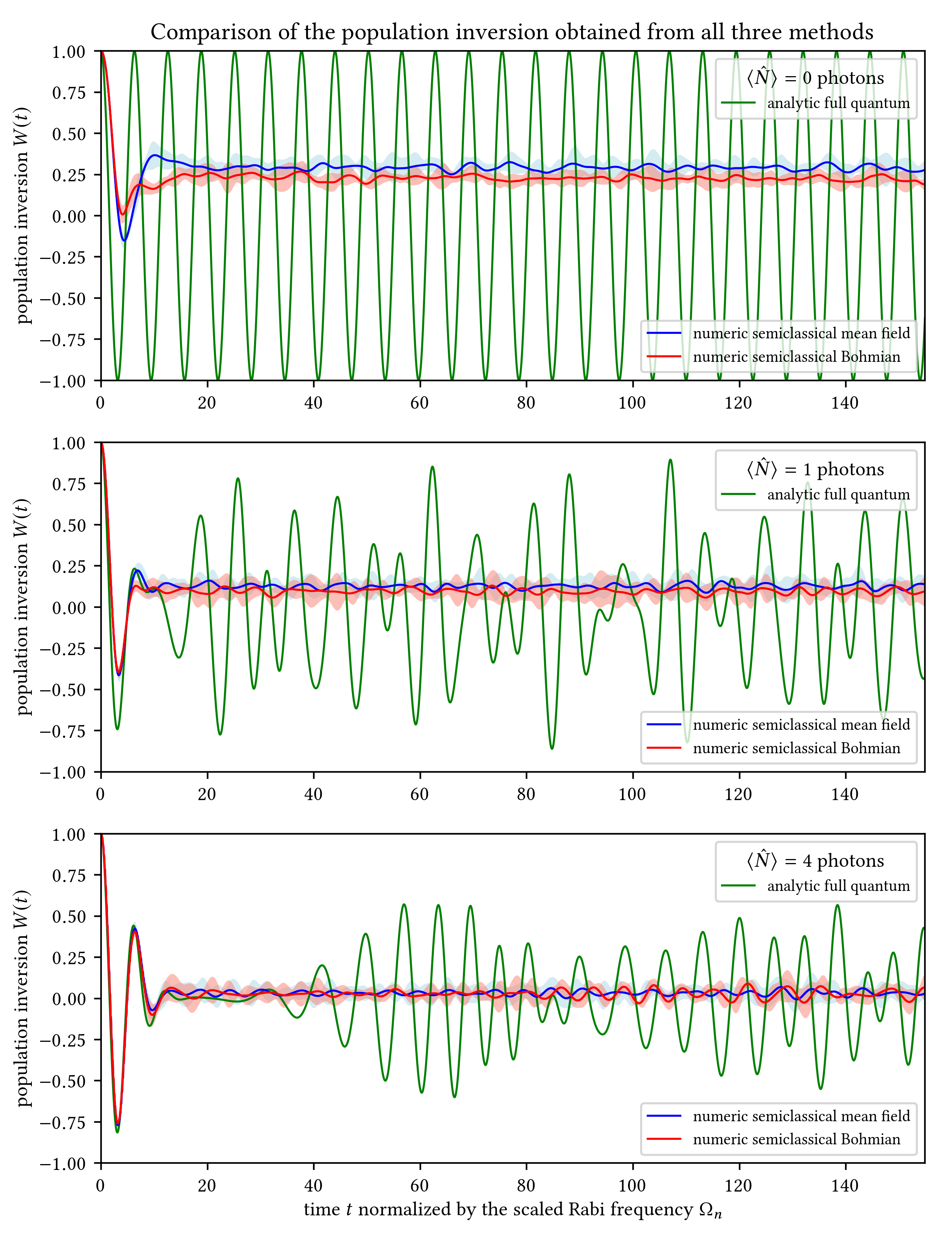}
        \caption{First part of the comparison between full quantum and semi-classical solutions in the rotating wave approximation for different expected photon numbers, here 0, 1 and 4.
        Confidence intervals of 2-$\sigma$ for the numerical data.}
        \label{fig:jaynes_cummings_comparison1}
\end{figure}

\begin{figure}[p]
        \centering
        \includegraphics[width=\textwidth]{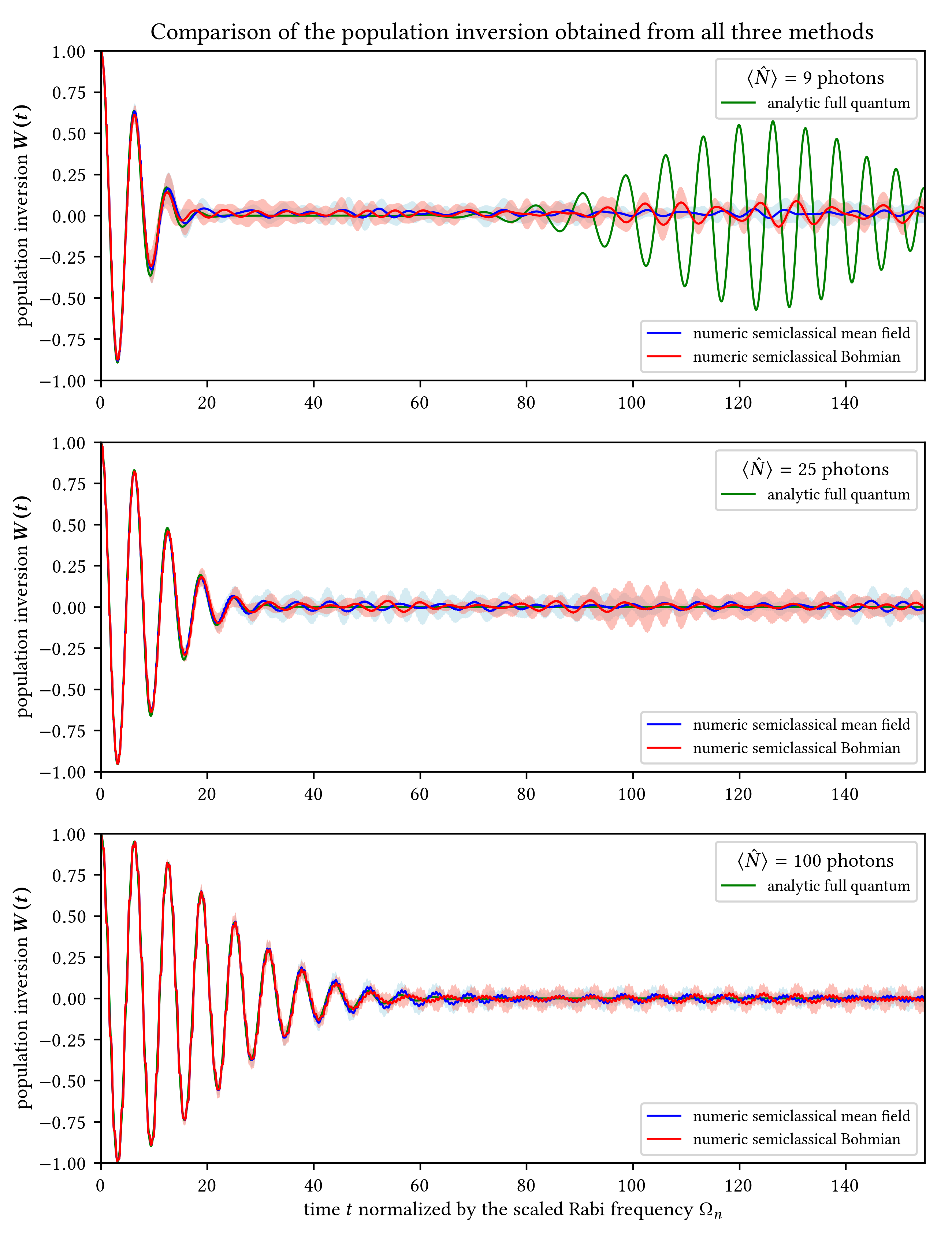}
        \caption{Second part of the comparison between full quantum and semi-classical solutions in the rotating wave approximation for different expected photon numbers, here 9, 25 and 100.
        Confidence intervals of 2-$\sigma$ for the numerical data.}
        \label{fig:jaynes_cummings_comparison2}
\end{figure}

Taking the 1s and 2p states, the parameters are given in table \ref{tab:jaynes_cummings_comparison}.
The choice of $\al$ ensures that we are in the weak coupling regime where the rotating wave approximation is applicable.
The full quantum result is obtained analytically using this approximation.
The resulting population inversion is shown in figures \ref{fig:jaynes_cummings_comparison1} and \ref{fig:jaynes_cummings_comparison2} for different expected photon numbers $\la {\widehat N} \ra$.
As these expected photon numbers are also considered in \cite{shore93}, they serve as a benchmark for our simulations, too.

\begin{table}[t]
        \centering
        \begin{tabular}{cc}
                \toprule
                \multicolumn{2}{c}{chosen parameters}\\
                \midrule
                $n_-$, $l_-$, $m_-$&1, 0, 0\\
                $n_+$, $l_+$, $m_+$&2, 1, 0\\
                $\alpha$&0.005\\
                ${\bf e}_p$&z-direction\\
                $\gamma$&0, 1, 2, 3, 5, 10\\
                \bottomrule
        \end{tabular}
        \hspace{1cm}
        \begin{tabular}{cc}
                \toprule
                \multicolumn{2}{c}{derived parameters}\\
                \midrule
                $\mathcal P$&$\approx0.745$\\
                $\nu$&0.375\\
                $g$&$\approx 0.0043$\\
        $\langle\widehat N\rangle$&0, 1, 4, 9, 25, 100\\
                \bottomrule
        \end{tabular}
        \caption{Choice of parameters for producing figures \ref{fig:jaynes_cummings_comparison1} and \ref{fig:jaynes_cummings_comparison2}. All numbers are given in atomic units.}
        \label{tab:jaynes_cummings_comparison}
\end{table}

\begin{figure}[t]
        \centering
        \includegraphics[width=\textwidth]{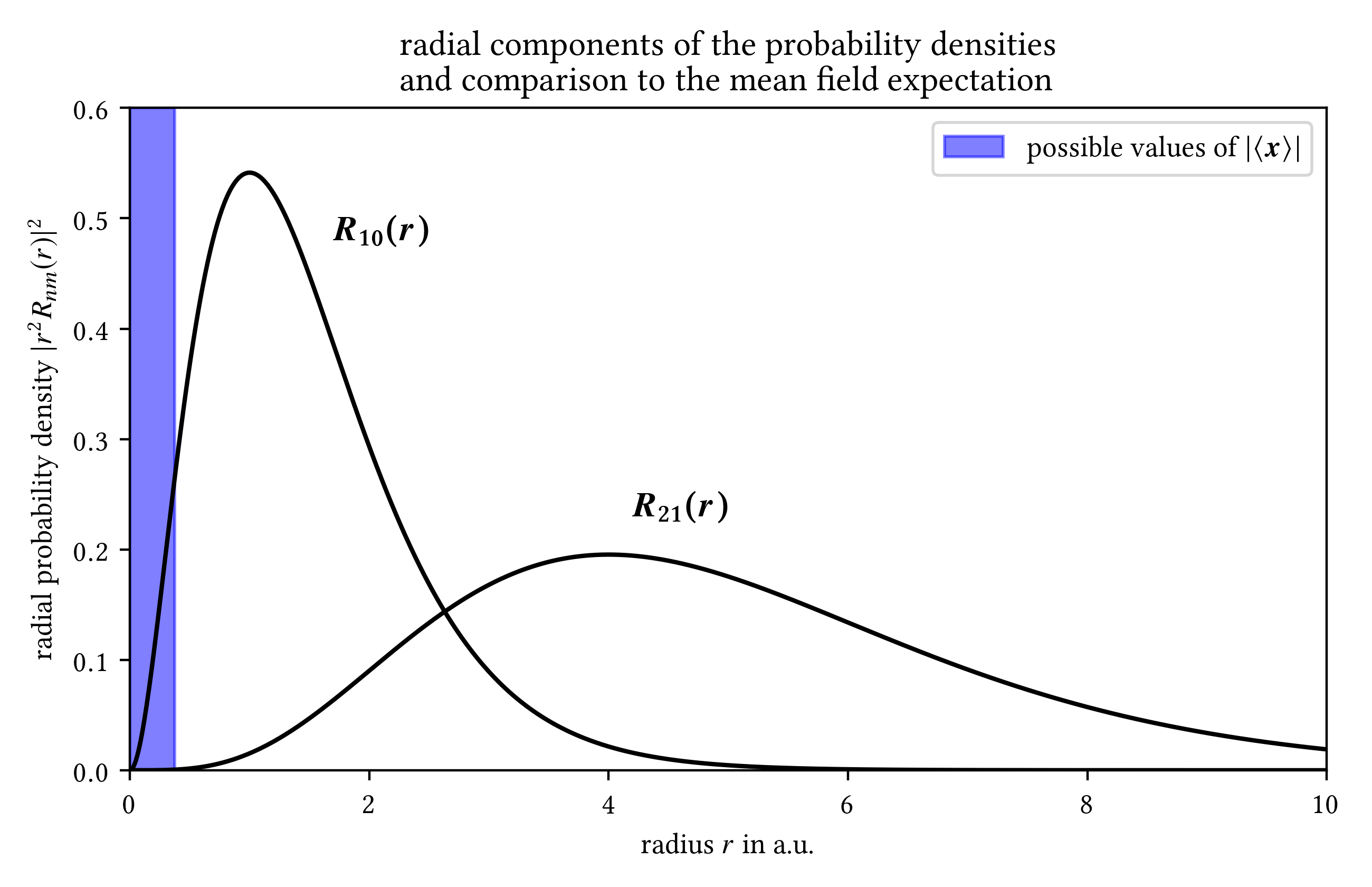}
        \caption{Visualization of the difference between backreaction terms.
                Radial probability densities of the 1s and 2p states.
                While the Bohmian back-reaction depends on ${\bf X} \cdot {\bf e}_p$, the mean field back-reaction depends on  $|\langle {\bf x} \rangle|\cdot {\bf e}_p$.
                The range of $|\langle {\bf x} \rangle|$ is $[0,{\mathcal P}/2]$, with ${\mathcal P}$ given in table \ref{tab:jaynes_cummings_comparison}, and is indicated in blue in the figure.
        Hence, the maximum absolute value of the Bohmian backreaction term is about an order of magnitude bigger than the mean field backreaction one.}
        \label{fig:radialdist}
\end{figure}


The mean field and Bohmian semi-classical approximation yield similar results, with confidence intervals that mostly overlap for each $\langle\widehat N\rangle$. 
The main difference among them actually arises from the different sets of initial field mode samples, which were sampled independently. This was checked by taking identical initial field configurations in each method. In that case (not shown), the two semi-classical graphs are essentially identical with differences only becoming visible upon sufficiently strong magnification of the plots.

At first sight, this may be surprising since the form of the backreaction from the quantum particle onto the field is very different.
In particular, while the Bohmian back-reaction depends on $ {\bf X} \cdot {\bf e}_p$, the mean field back-reaction depends on  $|\langle {\bf x} \rangle|\cdot {\bf e}_p$.
The difference between ${\bf X} \cdot {\bf e}_p$ and $|\langle {\bf x} \rangle|\cdot {\bf e}_p$ can be up to about one order of magnitude, as is illustrated in figure \ref{fig:radialdist}.
However, there is agreement between the two approaches because the weak coupling causes the backreaction to become negligible in both cases. In the next section, we will consider stronger couplings, which will require to go beyond the regime of the rotating wave approximation.

The plots show that the semi-classical approaches describe the collapse of the  Rabi oscillations in the full quantum result very well when the initial number of photons present in the cavity is large enough. This is as expected as the coherent state itself behaves more and more classical with increasing $\langle N \rangle$.
The mechanism which leads to the collapse in the semi-classical case is that each single realization of the system performs its Rabi oscillations with a slightly different frequency.
Initially, the phases are still correlated, but become decorrelated rather quickly.
Averaging the population inversion over the whole ensemble then leads to the collapse phenomenon.
What both the Bohmian and mean field approximation fail to reproduce are the genuine quantum effects like the vacuum Rabi oscillations and the revivals after collapse. 

From about $\langle \widehat N\rangle=4$ onward, the collapse is approximated very well by both semi-classical methods.
For $\langle \widehat N\rangle=1$, the semi-classical solutions quickly differ quantitatively and later also qualitatively from the quantum solution, which displays a somewhat erratic behavior.
In the $\langle \widehat N\rangle=0$ case, instead of Rabi oscillations, the population inversion rapidly drops down.
In all other plots, after the collapse, the semi-classical graphs drop down to about zero population inversion, which means that the atom is in the ground or first excited state with equal probability.
There are still small, irregular residual oscillations, which are essentially a consequence of the random sampling of initial conditions and they become smaller with increasing sample size.

\subsection{Transitions between 1s and 2p state, beyond weak coupling}
\begin{figure}[p]
        \centering
        \includegraphics[width=\textwidth]{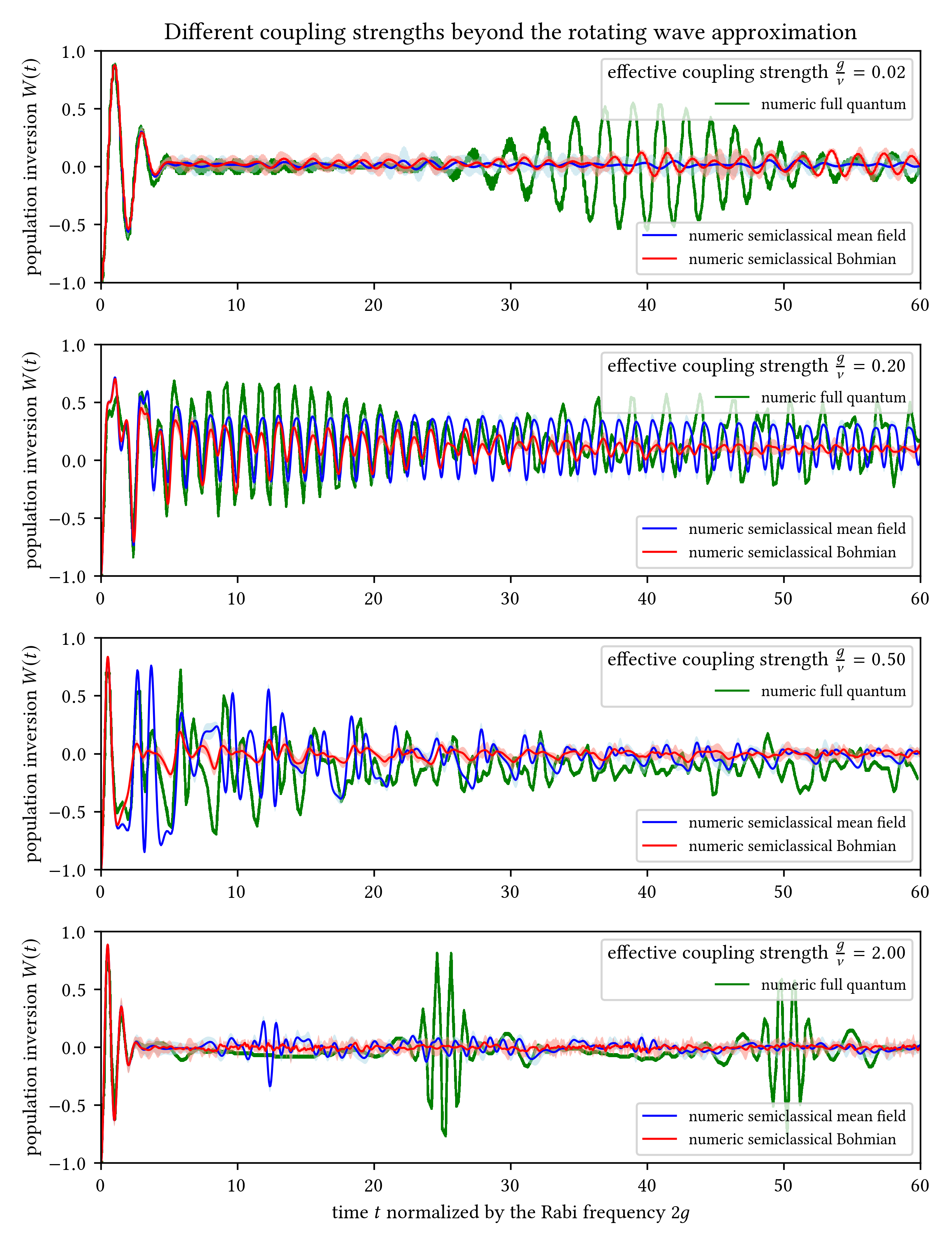}
        \caption{Direct comparison of the semi-classical results to the numerical data beyond the rotating wave approximation for a series of effective couplings.
                The green curves have been computed by Zhang, Chen and Zhu \cite[fig.~3]{zhang13}.
        With kind permission of Yu-Yu Zhang and Chinese Physics Letters.}
        \label{fig:beyond_rwa}
\end{figure}

We can extend the comparison beyond weak coupling, where the rotating phase approximation is no longer valid.
In \cite{zhang13}, the population inversion was numerically calculated for our system with stronger couplings \footnote{This means that instead of the usual Jaynes--Cummings Hamiltionian interaction term \eqref{eq:jaynes_cummings_interaction}, the full quantum 2-level Hamiltonian with the interaction term given in \eqref{eq:two_level_hamiltonian_terms}, for which no explicit solution is known, was solved numerically.}.
For the atom starting in the ground state and $\langle \widehat N\rangle=10$, their data is compared to the semi-classical approximations in figure \ref{fig:beyond_rwa}. 

Starting at a coupling $g=0.0075$ ($g/\nu = 0.02$) in the top plot, the regime where the rotating wave approximation becomes invalid is entered.
Both semi-classical approaches again reproduce the collapse well, but directly after that, the full quantum behavior displays some additional oscillations of very high frequency, which is absent in both semi-classical treatments.

Increasing the coupling to $g=0.075$ ($g/\nu = 0.2$) makes the revival phenomenon disappear in the full quantum results.
It gets replaced by an again very erratic behavior, which however still differs from the middle plot of figure \ref{fig:jaynes_cummings_comparison1}.
Both semi-classical graphs qualitatively follow the quantum solution up to $2gt\approx5$.
After that, all three curves separate.
The mean field semi-classical approximation seems slightly closer to the exact quantum result because of the larger amplitude of the oscillation, but the overall differences are still significant.

At $g=0.1875$ ($g/\nu = 0.5$), where the Rabi frequency matches the frequency of the counter-rotating terms, the curves becomes even more erratic.
Despite quantitatively being completely off, the mean field method at least qualitatively follows the quantum result up to about the same time as before.
The Bohmian one on the other hand rapidly collapses down around $2gt\approx3$ already and then settles to small amplitude oscillations around zero, which eventually die off. 

Finally, the collapse and revival phenomenon returns in the quantum description for extremely strong coupling $g=0.75$ ($g/\nu = 2$).
As has been the case in the weak coupling regime before, the semi-classical approximations accurately reproduce the collapse, but none of the revivals.
In the mean field case, at half of the first revival time, there is a small, but statistically significant surge in oscillation amplitude, resulting in what looks like a small version of the first revival event.

The overall conclusion is that in this case there are clear differences between the mean field and the Bohmian semi-classical approximations.
However, neither stands out in approximating the exact quantum result very well.
Again, the genuinely quantum effect of revival is never reproduced by any of the semi-classical approaches, while the collapse tends to be reproduced rather well, as is expected from the theory.

\subsection{Transitions between 1s and 9p state}\label{1s9p}
There is another way in which a quantitative difference between the two semi-classical approximations can be demonstrated without sacrificing the rotating wave approximation and its analytic full quantum result.
This is done by considering transitions between the ground state and a highly excited state, which we take to be the 9p state. In the case of the mean field semi-classical approximation, the backreaction from the atom onto the field will tend to be very small with ${\mathcal P} \approx 0.047$, because the energy eigenstates have little spatial overlap. On the other hand, in the Bohmian case, the magnitude of the backreaction gets boosted because the position ${\bf X}$ of the particle may be far away from the origin, so that $|{\bf X}|$ can be up to 250. In this way, the difference in the strength of the backreaction as visualized in figure \ref{fig:radialdist} is purposefully amplified to an extreme degree.

\begin{table}[t]
        \centering
        \begin{tabular}{cc}
                \toprule
                \multicolumn{2}{c}{fundamental parameters}\\
                \midrule
                $n_-$, $l_-$, $m_-$&1, 0, 0\\
                $n_+$, $l_+$, $m_+$&9, 1, 0\\
                $\alpha$&0.1\\
                ${\bf e}_p$&z-direction\\
                $\gamma$&10\\
                \bottomrule
        \end{tabular}
        \hspace{1cm}
        \begin{tabular}{cc}
                \toprule
                \multicolumn{2}{c}{derived parameters}\\
                \midrule
                $\mathcal P$&$\approx0.047$\\
                $\nu$&$\approx0.494$\\
                $g$&$\approx 0.005$\\
                $\langle \hat N\rangle$&100\\
                \bottomrule
        \end{tabular}
        \caption{Choice of parameters for producing figure \ref{fig:jaynes_cummings_excited}.
Left: parameters fixed in the model, right: values of constants as a consequence of the choice on the left.
All numbers given in atomic units.}
        \label{tab:jaynes_cummings_excited}
\end{table}
\begin{figure}[t]
        \centering
        \includegraphics[width=\textwidth]{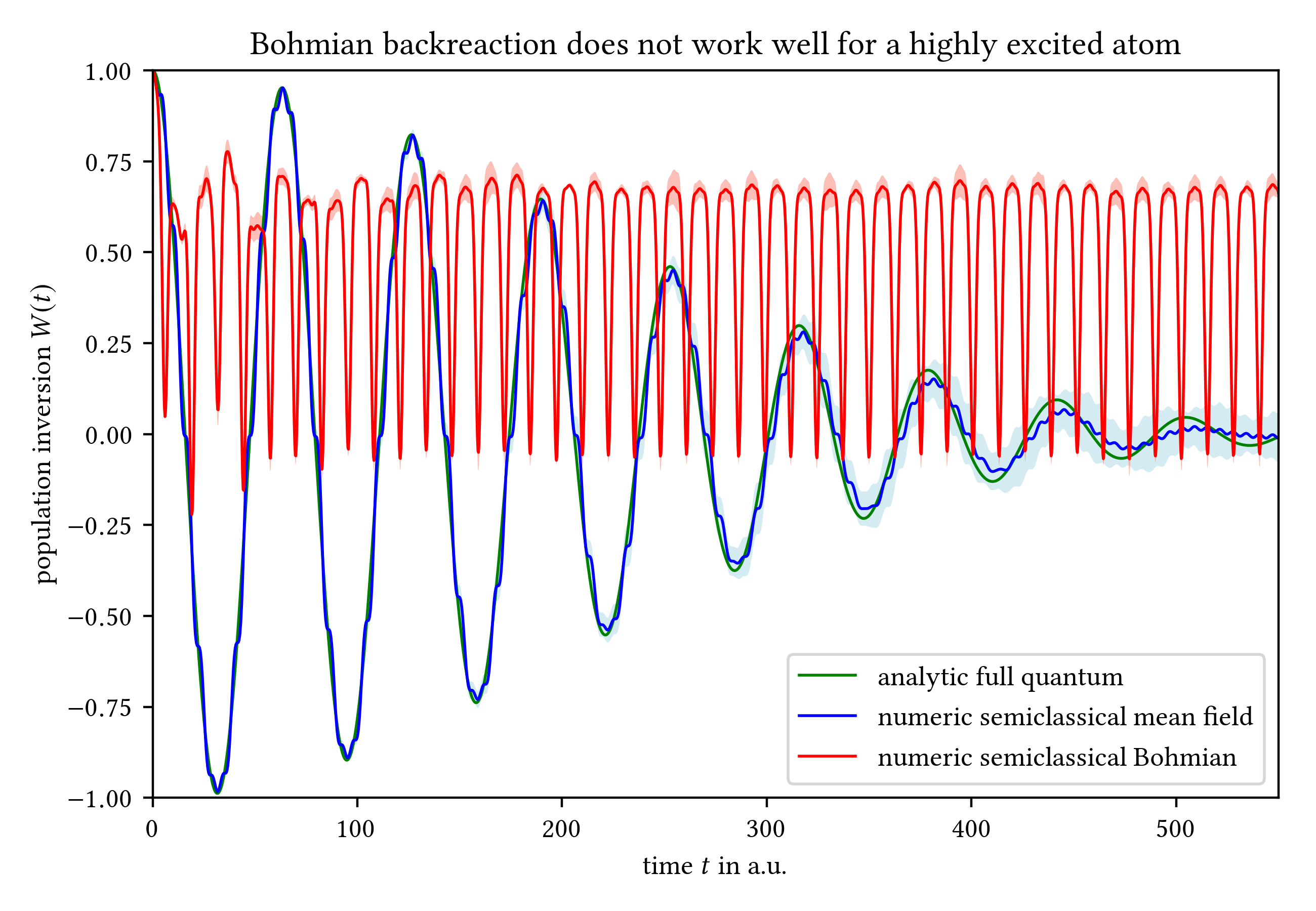}
        \caption{Comparison for a transition between ground state and a highly excited hydrogen eigenstate with $n=9$.}
        \label{fig:jaynes_cummings_excited}
\end{figure}

The full set of parameters is given in table \ref{tab:jaynes_cummings_excited}. We again use the rotating phase approximation to calculate the full quantum result. 
The results are given in figure \ref{fig:jaynes_cummings_excited}.
Only the initial phase of collapse is presented, in which there already is a clear difference among the two approximation methods when the atom again initially starts from the excited state.

The mean field approximation agrees very well with the full quantum result, with only significant deviations given by the small wiggles in the mean field result.
This is a consequence of neglecting the counter-rotating terms in the rotating wave approximation for the full quantum result.
The very same effect has also been present in figures \ref{fig:jaynes_cummings_comparison1} and \ref{fig:jaynes_cummings_comparison2}, however less prominent.
The Bohmian semi-classical approximation, however, differs a lot from the full quantum result.
It features pronounced oscillations apparently with the frequency of the external field, although no longer around $W=0$, but heavily skewed towards the excited state.
In the excited state, many particles have large values of X, thus creating strong fields which couple back to the quantum system apparently protecting it from falling to the ground state.
In contrast, the magnitude of the backreaction is negligible in the mean field case, which explains why the usual semiclassical approximation reproduces the collapse of population inversion oscillations well.
In fact, the absence of a backreaction term, i.e., putting the right hand side of equations \eqref{101} or \eqref{eq:semiclassical_classical_part_Bohmian} to zero, yields a very similar curve.

Note that such a discrepancy between the two approaches is observed exclusively when the atom initially is in the excited state.
When taking the ground state as initial value for the population inversion and sampling the Bohmian particle's initial position from that state, there is again no discernible difference between the two semi-classical approaches (not shown).

In the full quantum theory, the position of a Bohmian particle tends to follow the bulk of the packet, because of equivariance.
So in particular, if the state is near the ground state or near the excited state, then the particle will (usually) be in the respective support.
However, the particle no longer follows the bulk of the packet in the case of the Bohmian semi-classical approximation, because of the lack of equivariance.
Even though the wave function might transition from the excited state to the ground state, the particle tends not to move to the support of the ground state. It seems to be obstructed from doing so because of the radial nodes that emerge dynamically. This was observed when analyzing trajectories of such particles.

As mentioned before in section \ref{bsa}, there is an issue concerning the definition of the population inversion.
In the case of non-overlapping energy eigenstates it would make more sense to define it in terms of the particle positions rather than the wave function.
But with particles not being able to move between eigenstates in the Bohmian semi-classical approximation, such a definition would not bring the results of this method closer to the exact quantum results.
Again the population inversion will not go to zero.  

\section{Conclusion}\label{conclusion}
We have compared the usual mean field semi-classical approximation with the Bohmian one in the case of the Rabi model, which poses a practical scenario for studying matter-light interaction. 
The full quantum treatment of the Rabi model displays distinct features coming from the quantum description of the electric field, which is given by the revival of the population inversion.
Such a revival was not produced by either semi-classical approximation.
On the other hand, both semi-classical approximations reproduced the collapse of the population inversion rather well.
Overall, the mean field semi-classical approximation and the Bohmian one gave comparable results.

The reason that the Bohmian semi-classical approximation did not give better results seems to be due to the particular form of interaction.
The improvement reported in for example \cite{prezhdo01} was due to the fact that the Bohmian approximation preserved the linearity of the full quantum theory when the wave function is in a superposition of wave functions with disjoint support.
Because of the particular type of interaction considered here, the Bohmian approximation does not preserve that linearity.

Only the lowest order semi-classical approximation was considered here.
Improvement is expected to be obtained by considering higher order corrections.
A way to obtain such corrections was outlined in \cite{oriols07,norsen15}.
It would be interesting to see whether revival of the population inversion shows up in such higher order corrections.

It is important to have tests of the Bohmian semi-classical approximation in cases where the exact quantum result is known because it gives clues about the domain of validity and applicability where the exact quantum result is not known, like in for example quantum gravity.

\section{Acknowledgments}
It is a pleasure to thank Pablo Barberis-Blostein for suggesting this project.
WS was supported by the Deutsche Forschungsgemeinschaft. Currently, WS is
supported by the Research Foundation Flanders (Fonds Wetenschappelijk
Onderzoek, FWO), Grant No. G066918N. The work of LK and DAD was supported by
the junior research group ``Interaction between Light and Matter'' of the Elite
Network of Bavaria.

\end{document}